\documentclass[twocolumn,showpacs,preprintnumbers,superscriptaddress]{revtex4}
\usepackage{amsmath}
\usepackage{dcolumn}
\usepackage{ifpdf}
\usepackage{bm}
\usepackage{epsfig}
\usepackage{graphicx}
\usepackage{amsfonts}
\usepackage{amssymb}
\usepackage{mathrsfs}
\usepackage{color}

\begin{document}

\title{Precision measurement of electrical charge with optomechanically
induced transparency}
\author{Jian-Qi Zhang}
\affiliation{State Key Laboratory of Magnetic Resonance and Atomic and Molecular Physics,
Wuhan Institute of Physics and Mathematics, Chinese Academy of Sciences,
Wuhan 430071, China }
\author{Yong Li}
\email[Corresponding author Email: ]{liyong@csrc.ac.cn}
\affiliation{Beijing Computational Science Research Center, Beijing 100084, China }
\author{Mang Feng}
\email[Corresponding author Email: ]{mangfeng@wipm.ac.cn}
\affiliation{State Key Laboratory of Magnetic Resonance and Atomic and Molecular Physics,
Wuhan Institute of Physics and Mathematics, Chinese Academy of Sciences,
Wuhan 430071, China }
\author{Yi Xu}
\affiliation{Laboratory of Photonic Information Technology, School for Information and
Optoelectronic Science and Engineering, South China Normal University,
Guangzhou 510006, China}
\affiliation{School of Physics and Electric Engineering, Guangzhou University, Guangzhou
510006, China}

\begin{abstract}
We propose a potentially practical scheme to precisely measure the charge
number of small charged objects by using optomechanically induced
transparency (OMIT) in optomechanical systems. In contrast to conventional
measurements based on noise backaction on optomechanical systems, our scheme
presents an alternative way to detect the charge number exactly, by
monitoring small deformation of the mechanical resonator sensitive to the
charge number of nearby charged object. The relationship between the charge
number and the OMIT window width is investigated and the feasibility of the
scheme is justified by numerical simulation with currently available
experimental values.
\end{abstract}

\date{\today }
\maketitle


\section{introduction}

Precision measurement is one of the essential tasks in the study of modern
physics. The (micro- or nano-) mechanical resonators (MRs) hold the promise
for realizing precision measurements due to the possibility of presenting
both classical and quantum properties~\cite{V.B.Braginsky-Measurements of
Weak Forces in Physics Experiments,prl-97-237201}. For measurements
approaching quantum limit~\cite{Physics-2-40} with mechanical systems, we
have to cool the MRs to their ground states and show obvious quantum
behavior. Up to now, various methods for MR cooling have been proposed in
optomechanical and electromechanical systems, such as feedback cooling \cite%
{PRB-78-134301, Nature-460-724, Nature-432-200}, backaction sideband cooling
\cite{prl-92-075507, prl-99-093901, prl-99-093902}, bang-bang cooling~\cite%
{zhang-PRL}, electromagnetically induced transparency cooling \cite%
{prl-103-227203}, measurement-based cooling \cite{PRB-84-094502}, and
thermal light cooling \cite{prl-108-120602}, some of which have been
achieved experimentally \cite{Nature-432-1002, NatPhys-5-485,Nature-463-72,
Nature-475-359, Nature-478-89, Nature-464-697}.

The precision measurement based on MRs can be classified by two kinds of
systems, i.e., optomechanical and electromechanical systems. We focus in the
present work on the optomechanical system, in which the precision
measurements were usually carried out via the correlations between the
output spectra and measured quantities based on reflected noise \cite%
{Physics-2-40}. For example, precision measurement of displacement of the MR
has been achieved with a factor of five times higher than standard quantum
limit in optical output spectra \cite{NatPhys-5-509}, and a recent
experimental report was published for displacement measurement of the MR
beyond standard quantum limit \cite{prl-104-133602}.

The optomechanically induced transparency (OMIT) is a kind of induced
transparency caused by radiation pressure to couple light to MR modes \cite%
{Science-330-1520}. Recently, the OMIT in optomechanical system has been
predicted theoretically \cite{pra-81-041803, Science-330-1520,pra-86-013815}
and also observed experimentally \cite{Science-330-1520, NatPhoton-4-236,
Nature-471-204, Nature-472-69}. However, the application of OMIT has not yet
been fully explored. As far as we know, the proposed applications are only
for slow light with OMIT controlling the speed of light \cite{Nature-472-69}
and for single photon router with OMIT to control the probe field in a
single photon Fock state \cite{pra-85-021801}.

The aim of the present work is to detect the charge number of a small
charged body via OMIT in an opto-mechanical and electrical system. The
Coulomb interaction between a charged MR and a nearby charged body in such a
hybrid system will modify both the steady-state position of the MR and the
mean photon number in the cavity, which affect significantly the window
width of the OMIT \cite{pra-81-041803,Science-330-1520}. As a result, the
charge number of the charged object is possibly detected via monitoring the
modified window width of the OMIT.

Our study shows that the window widths in some special regions of the OMIT
vary with the charge number in a sensitive way, which makes it possible for
a precision measurement of the charge number. We notice that previous ideas
for ultra-sensitive measurements in opto-mechanical systems, e.g., cavity
optomechanical magnetometer \cite{prl-108-120801} and displacement
measurement \cite{NatPhys-5-509}, are based on quantum noise backaction \cite%
{Physics-2-40}. In contrast, the noise backaction is unnecessary in our case
because the output intensity of the probe field is monitored via the OMIT
based on the expectation of massive photons. Moreover, conventional MR
electrometers, such as vibrating reed electrometers \cite{small-1-786}, are
formed by movable and fixed electrodes, which can be used to measure the
Coulomb forces with variable capacitors. Limited by the extremely sensitive
electricity, e.g., $0.12$ aA current \cite{IEEE}, however, the MR
electrometers cannot measure the charge densities in tiny objects (e.g., $<6$
nm \cite{J-Phys-Conf-Ser-304-012064}). In contrast, since optical
measurements are usually more sensitive than electrical ones, our scheme in
an optical way should work well even for detecting single charges in very
small objects. Furthermore, our scheme makes use of the unique feature of
optomechanical measurements which is considered to have higher sensitivity
than electromechanical measurements \cite{apl-100-171111} .

Our system consisting of both an optomechanical system and a charged object
is actually a cavity opto-electromechanical system, belonging to a currently
active research area. For example, a recent idea for detecting electric
gradient force was proposed by using a net dipole moment in a microtoroid
\cite{prl-104-123604,pra-82-023825}. In comparison, our scheme using OMIT
can be applied to measure weaker force since the net charge density on a
metal surface can be much higher than the net dipole moment density in a
semiconduction surface. Moreover, there was a proposal to measure
displacements and forces by noise spectra \cite{njp-14-075015}. Honestly
speaking, our scheme based on the expectation of massive photons might not
work better than in Ref. \cite{njp-14-075015} (Concrete estimates will be
given later in the section of Conclusion). But our focus is on the electric
charge detection, which is a practical application of OMIT.

The paper is structured as follows: We present the model and Hamiltonian of
the system in next section, and study the output field for the OMIT in Sec.
III. The relationship between the charge number and the output field as well
as the feasibility of our scheme is described in Sec. IV. The last section
is for a brief conclusion.

\section{Model and Hamiltonian}

\begin{figure}[tbph]
\includegraphics[width=8 cm]{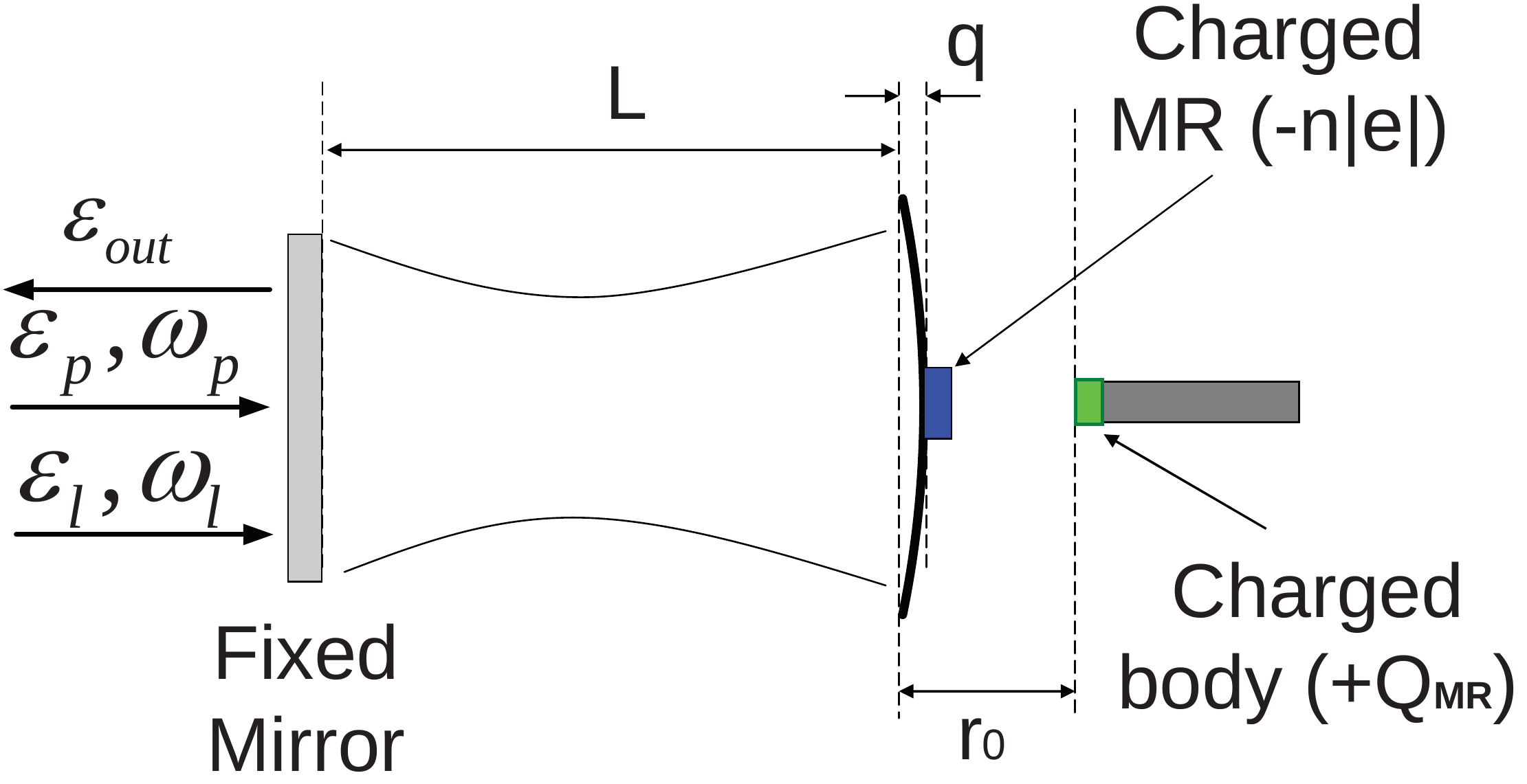}
\caption{Schematic diagram of the system. An optomechanical cavity with the
length $L$ is driven by two light fields. One is the pumping field $\protect%
\varepsilon_{l}$ with frequency $\protect\omega_{l}$ and the other is the
probe field $\protect\varepsilon_{p}$ with frequency $\protect\omega_{p}$.
The output field is represented by $\protect\varepsilon_{out}$. $r_{0}$ is
the distance between the charged body and the charged MR in absence of the
radiation pressure and the Coulomb force. Under the action of the radiation
pressure and the Coulomb force, the MR takes a position $q$. Here, the
charges on the charged body and charged MR are $-n|e|$ and $Q_{MR}$,
respectively.}
\label{fig 1}
\end{figure}

The model we consider is sketched in Fig.~\ref{fig 1}, where a high-quality
cavity consists of a fixed mirror and a movable one, i.e., a MR. Besides the
radiation pressure force coupling the MR to the cavity mode, the charged MR
is subject to the Coulomb force due to the charged body nearby. Such a
system can be described as
\begin{equation}
\begin{array}{lll}
H_{1} & = & \hbar \omega _{c}c^{\dag }c+(\dfrac{p^{2}}{2m}+\dfrac{m\omega
_{m}^{2}}{2}q^{2})-\chi qc^{\dag }c \\
& - & \dfrac{n|e|Q_{MR}}{4\pi \epsilon (r_{0}-q)}+i\hbar \lbrack
(\varepsilon _{l}e^{-i\omega _{l}t}+\varepsilon _{p}e^{-i\omega
_{p}t})c^{\dag }-\text{H.c.}].%
\end{array}
\label{Q1-0}
\end{equation}
Here the first term is for the single-mode cavity of eigen-frequency $\omega
_{c}$ with the bosonic annihilation (creation) operator $c$ ($c^{\dagger}$).
The second term describes the vibration of the MR where $q$ and $p$ are,
respectively, the position and momentum operators of the MR with
eigen-frequency $\omega _{m}$ and effective mass $m$. The third term is for
the radiation pressure coupling the cavity field to the MR, with $\chi
=\hbar \omega _{c}/L$ the coupling strength and $L$ being the cavity length.
The forth term presents the interaction of the charged MR with the charged
body via a Coulomb potential $V_{c}=\dfrac{-n|e|Q_{MR}}{4\pi \epsilon
(r_{0}-q)}$, where $Q_{MR} $ is the positive charge on the MR, $-n|e|$ is
for $n$ negative charges of the charged body to be detected, and $r_{0}$ is
the distance between the equilibrium positions of the MR center-of-mass and
the charged body in absence of the radiation pressure and the Coulomb force.
In our case with the attractive force, the Coulomb force on the MR points to
the same direction as the radiation pressure force on the MR. The last term
in Eq.~(\ref{Q1-0}) describes two optical drives to the cavity from the
fixed mirror: One is the strong pumping field with frequency $\omega _{l}$
and the other is the weak probe field with frequency $\omega _{p}$, and $%
\varepsilon _{l}$ and $\varepsilon _{p}$ are the corresponding driving
strengths, respectively.

In the case of $q\ll r_{0}$, the Coulomb interaction can be rewritten as $%
V_{c}\simeq -\dfrac{n|e|Q_{MR}}{4\pi \epsilon r_{0}}(1+\dfrac{q}{r_{0}})$.
Omitting the constant term, we have the Hamiltonian in the frame rotating
with the driving frequency $\omega _{l}$,
\begin{equation}
\begin{array}{lll}
H_{2} & = & \hbar \Delta _{c}c^{\dag }c+\dfrac{1}{2m}({p^{2}}+{m^{2}\omega
_{m}^{2}}q^{2})-\chi qc^{\dag }c \\
& - & n\eta q+i\hbar \lbrack (\varepsilon _{l}+\varepsilon _{p}e^{-i\delta
t})c^{\dag }-\text{H.c.}],%
\end{array}
\label{Q1-2}
\end{equation}
where $\Delta _{c}=\omega _{c}-\omega _{l}$, $\eta =|e|Q_{MR}/(4\pi \epsilon
r_{0}^{2})$, and $\delta =\omega _{p}-\omega _{l}$. Here both $\varepsilon
_{l}$ and $\varepsilon _{p}$ are complex.

\section{Mean-value equations and quadratures of the output field}

For analyzing the mean response of the OMIT, we may consider the Langevin
equations \cite{pra-83-023823} by neglecting quantum fluctuation of the
system \cite{pra-81-041803}. In our case, the mean-value equations of the
system are written as,
\begin{equation}
\begin{array}{lll}
\left\langle \dfrac{dq}{dt}\right\rangle & = & \dfrac{\left\langle
p\right\rangle }{m}, \\
\left\langle \dfrac{dp}{dt}\right\rangle & = & -m\omega _{m}^{2}\left\langle
q\right\rangle +n\eta +\chi \left\langle c^{\dag }c\right\rangle -\gamma
_{m}\left\langle p\right\rangle , \\
\left\langle \dfrac{dc}{dt}\right\rangle & = & -[\kappa +i(\Delta _{c}-%
\dfrac{\chi }{\hbar }\left\langle q\right\rangle )]\left\langle
c\right\rangle +\varepsilon _{l}+\varepsilon _{p}e^{-i\delta t},%
\end{array}
\label{Q2-1}
\end{equation}
where $\kappa $ and $\gamma _{m}$ are introduced as the decay rates of the
cavity and the MR, respectively. Eq. (\ref{Q2-1}) can be solved under the
condition that the pumping field is much stronger than the probe one. Eq.~(%
\ref{Q2-1}) is a nonlinear equation and the steady-state response in the
frequency domain is composed of many frequency components. We suppose the
steady-state solutions to Eq. (\ref{Q2-1}) taking the form of
\begin{equation}
\begin{array}{lll}
\left\langle q\right\rangle & = & p_{s}+p_{+}\varepsilon _{p}e^{-i\delta
t}+p_{-}\varepsilon _{p}^{\ast }e^{i\delta t}, \\
\left\langle p\right\rangle & = & q_{s}+q_{+}\varepsilon _{p}e^{-i\delta
t}+q_{-}\varepsilon _{p}^{\ast }e^{i\delta t}, \\
\left\langle c\right\rangle & = & c_{s}+c_{+}\varepsilon _{p}e^{-i\delta
t}+c_{-}\varepsilon _{p}^{\ast }e^{i\delta t},%
\end{array}
\label{Q2-2}
\end{equation}
where each solution contains three items $O_{s}$, $O_{+}$ and $O_{-}$ (with $%
O=p,q,c$), corresponding to the responses at the original frequencies $%
\omega _{l}$, $\omega _{p}$, and $2\omega _{l}-\omega _{p}$, respectively
\cite{pra-83-023823}. Since $O_{s}\gg O_{\pm }$, Eq. (\ref{Q2-2}) can be
solved by treating $O_{\pm }$ as perturbation. After combining Eq. (\ref%
{Q2-2}) with Eq. (\ref{Q2-1}), and ignoring the second-order small terms, we
obtain the steady-state mean-values of the system by resorting the
prefactors in terms of the exponentials $e^{\pm i\delta t}$
\begin{equation}
\begin{array}{llllll}
p_{s} & = & 0, & q_{s} & = & \dfrac{\chi |c_{s}|^{2}+n\eta }{m\omega
_{m}^{2} }, \\
c_{s} & = & \dfrac{\varepsilon _{l}}{\kappa +i\Delta }, & |c_{s}|^{2} & = &
\dfrac{|\varepsilon _{l}|^{2}}{\kappa ^{2}+\Delta ^{2}},%
\end{array}
\label{Q2-3}
\end{equation}
with $\Delta =\Delta _{c}-\chi q_{s}/\hbar $, and the solution of $c_{+}$
\cite{pra-81-041803} is
\begin{equation}
\begin{array}{lll}
c_{+} & = & \dfrac{(\delta ^{2}-\omega _{m}^{2}+i\gamma _{m}\delta )[\kappa
-i(\Delta +\delta )]-2i\omega _{m}\beta }{[\Delta ^{2}+(\kappa -i\delta
)^{2}](\delta ^{2}-\omega _{m}^{2}+i\delta \gamma _{m})+4\Delta \omega
_{m}\beta }%
\end{array}
\label{Q2-41}
\end{equation}%
with $\beta =\chi ^{2}|c_{s}|^{2}/(2m\hbar \omega _{m})$.

Making use of the input-output relation of the cavity \cite%
{D.F.Walls-Quantum Optics}, we have the output field
\begin{equation}
\begin{array}{cll}
\varepsilon _{out} & = & \varepsilon _{in}-2\kappa c \\
& = & \varepsilon _{l}+\varepsilon _{p}e^{-i\delta t}-2\kappa
(c_{s}+c_{+}\varepsilon _{p}e^{-i\delta t}+c_{-}\varepsilon _{p}^{\ast
}e^{i\delta t}),%
\end{array}
\label{Q2-5}
\end{equation}
and thereby the transmission of the probe field is given by \cite%
{Science-330-1520}
\begin{equation}
\begin{array}{cll}
t_{p} & = & \dfrac{\varepsilon _{p}-2\kappa \varepsilon _{p}c_{+}}{
\varepsilon _{p}}=1-2\kappa c_{+},%
\end{array}
\label{Q2-55}
\end{equation}
which can be measured by homodyne technique \cite{D.F.Walls-Quantum Optics}.

Defining $\varepsilon _{T}=2\kappa c_{+}$, we obtain the quadrature $%
\varepsilon _{T}$ of the optical components with frequency $\omega _{p}$ in
the output field,
\begin{equation}
\begin{array}{cll}
\varepsilon _{T} & = & 2\kappa \dfrac{(\delta ^{2}-\omega _{m}^{2}+i\gamma
_{m}\delta )[\kappa -i(\Delta +\delta )]-2i\omega _{m}\beta }{[\Delta
^{2}+(\kappa -i\delta )^{2}](\delta ^{2}-\omega _{m}^{2}+i\delta \gamma
_{m})+4\Delta \omega _{m}\beta },%
\end{array}
\label{Q2-6}
\end{equation}
whose real and imaginary parts, Re$[\varepsilon _{T}]$ and Im$[\varepsilon
_{T}]$, represent the absorptive and dispersive behavior of the
optomechanical system, respectively \cite{pra-81-041803}.

In order to reduce Eq. (\ref{Q2-6}) and understand the relationship between
the charge number and the OMIT, we assume following conditions \cite%
{pra-81-041803, Science-330-1520}: (i) $\Delta \simeq \omega _{m}$ and (ii) $%
\omega_{m}\gg \kappa$. The first condition means the frequency of the cavity
to be resonant with that of the optomechanical anti-Stokes sideband, which
actually leads to optimal cooling. The second condition is the well-known
resolved sideband condition, which ensures the OMIT splitting to be
distinguished \cite{Science-330-1520}. Moreover, it is known that the
coupling between the MR and the cavity is strongest in the case of $\delta
\simeq \omega _{m}$ \cite{pra-81-041803}, which makes $\delta ^{2}-\omega
_{m}^{2}\simeq 2\omega_{m}(\delta -\omega_{m})$ achievable. Under these
conditions, we rewrite the output field as
\begin{equation}
\begin{array}{ccc}
\varepsilon _{T} & \approx & \dfrac{2\kappa }{\kappa -i(\delta -\omega _{m})+%
\dfrac{\beta }{\dfrac{\gamma _{m}}{2}-i(\delta -\omega _{m})}}.%
\end{array}
\label{Q2-7}
\end{equation}
Compared with the expression of the output field in Eq. (7) in Ref. \cite%
{pra-81-041803}, the parameter $\beta $ in Eq. (\ref{Q2-7}) in our case is
modified to be a function of the charge number $n$. Under appropriate
conditions, the window width of the OMIT can be used to identify the charge
number of the charged body.

\section{The charge number and the output field}

We show below in details how the charge number impacts the mean photon
number and how to measure the charge number by the window width of the OMIT.

From Eq. (\ref{Q2-3}), we have a third-order nonlinear equation for the MR
position $q_{s}$,
\begin{equation}
\begin{array}{l}
aq_{s}^{3}+bq_{s}^{2}+fq_{s}+d=0,%
\end{array}
\label{Q3-11}
\end{equation}
with
\begin{eqnarray}
a &=&m\omega _{m}^{2}\dfrac{\chi ^{2}}{\hbar ^{2}},  \notag \\
b &=&-2m\omega _{m}^{2}\dfrac{\chi }{\hbar }(\Delta _{c})-n\eta \dfrac{\chi
^{2}}{\hbar ^{2}},  \notag \\
f &=&m\omega _{m}^{2}\kappa ^{2}+m\omega _{m}^{2}(\Delta _{c})^{2}+2n\eta
(\Delta _{c})\dfrac{\chi }{\hbar },  \notag \\
d &=&-n\eta \kappa ^{2}-n\eta (\Delta _{c})^{2}-\chi |\varepsilon _{l}|^{2}.
\end{eqnarray}

\begin{figure}[tbph]
\includegraphics[width=8cm]{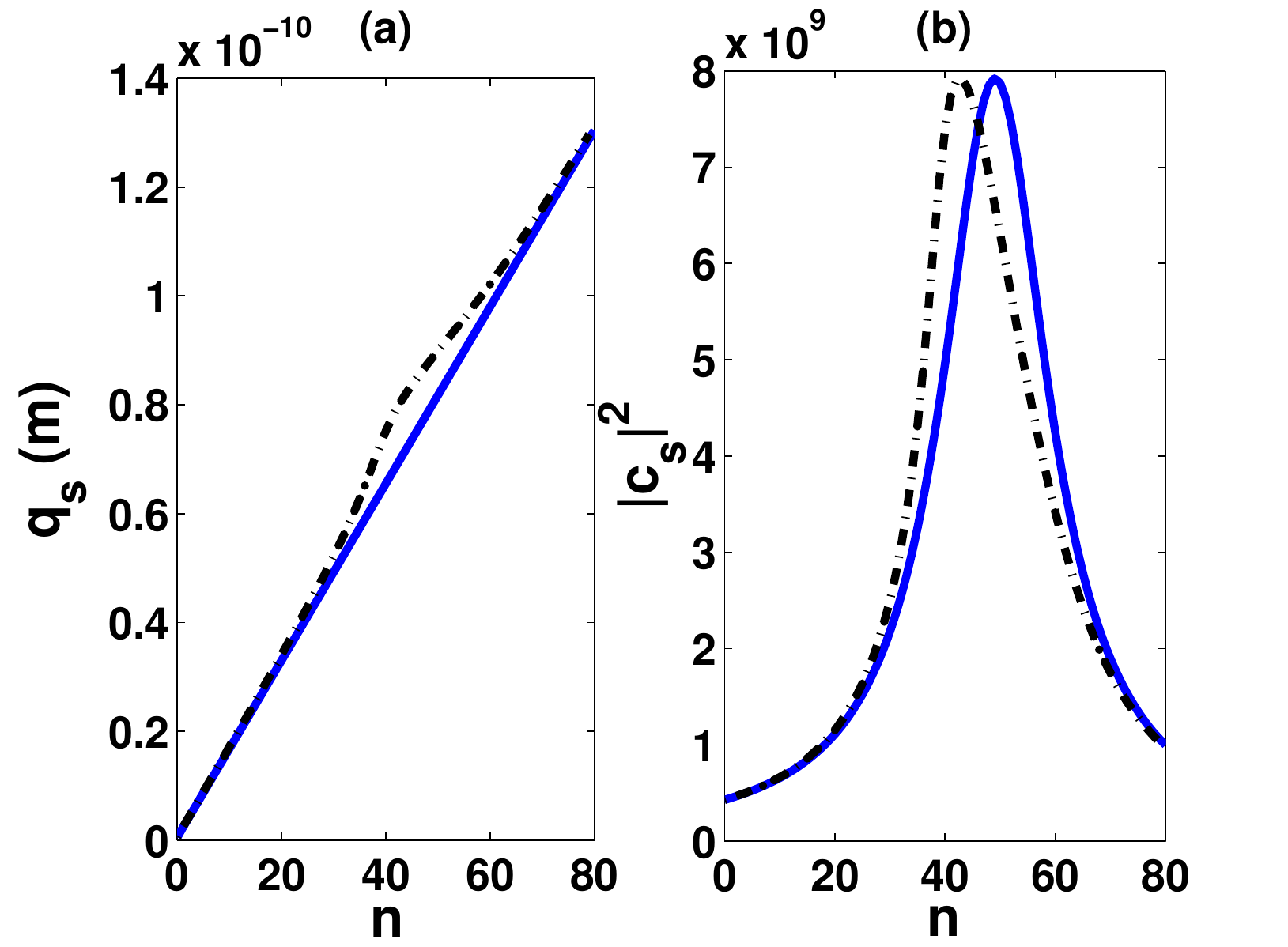}
\caption{(Color online) (a) The position $q_{s}$ versus the charge number $n$
; (b) The mean photon number as a function of the charge number $n$. The
black dotted-dashed lines [blue solid lines] represent the exact values [the
approximate values] of $q_{s}$ and $|c_{s}|^{2}$ using Eq. (\protect\ref%
{Q3-11}) and Eq. (\protect\ref{Q2-3}) [Eq. (\protect\ref{Q3-4}) and Eq. (
\protect\ref{Q3-5})], respectively. The values are taken from the
experiments in Refs. \protect\cite{pra-81-041803, Nature-460-724,
Science-304-74, pra-72-041405, Nature-432-200, NatPhys-6-707} as $k=8.897$N$%
\cdot $m$^{2}$/C$^{2}$ $\protect\lambda _{c}\equiv 2\protect\pi c/\protect%
\omega _{c}=1064$ nm, $L=25$ mm, $m=145$ ng, $\protect\kappa =2\protect\pi %
\times 215$ kHz, $\protect\omega _{m}=2\protect\pi \times 947$ kHz, $\protect%
\gamma _{m}=2\protect\pi \times 141$ Hz, $r_{0}=67$ $\protect\mu $m , $%
\protect\varepsilon _{l}=\protect\sqrt{2P\protect\kappa /\hbar \protect%
\omega _{c}}$ with $P=1$ mW, $Q_{MR}=CU$, $C=27.5$ nF and $U=1$ V.}
\label{fig21}
\end{figure}

To get a more intuitive understanding of the role that the Coulomb
interaction plays in Eqs. (\ref{Q2-6}) and (\ref{Q2-7}), we suppose $n\eta
\gg \chi |c_{s}|^{2}$, and obtain the solutions to Eq. (\ref{Q3-11}) as
\begin{equation}
q_{s}=\left\{
\begin{array}{cc}
\chi |c_{s0}|^{2}/(m\omega _{m}^{2}), & ~~~(n=0) \\
n\eta /(m\omega _{m}^{2}), & ~~~(n\geq 1)%
\end{array}%
\right.   \label{Q3-4}
\end{equation}%
with $|c_{s0}|^{2}$ being the mean photon number in absence of the Coulomb
interaction between the MR and charged object. The above equation means
that, for no charge in the system, the MR has a steady-state position $%
q_{s}=\chi |c_{s0}|^{2}/(m\omega _{m}^{2})$ under the action of pumping
field. In the presence of the attractive Coulomb interaction between the
charged body and the MR, the steady-state position of the MR is modified. In
the case of $n\eta \gg \chi |c_{s}|^{2}$, the steady-state position $%
q_{s}=n\eta /(m\omega _{m}^{2})$ reduces to a function of charge number in
the object. The phenomena have been shown in Fig. \ref{fig21}(a).

Moreover, from Eqs. (\ref{Q2-3}) and (\ref{Q3-4}), the mean photon number
takes the form of
\begin{equation}
|c_{s}|^{2}=\left\{
\begin{array}{cc}
|c_{s0}|^{2}, & ~~~~(n=0) \\
\dfrac{|\varepsilon _{l}|^{2}}{\kappa ^{2}+(\Delta _{c}-\dfrac{\chi }{\hbar }
\dfrac{n\eta }{m\omega _{m}^{2}})^{2}}, & ~~~~(n\geq 1)%
\end{array}%
\right.  \label{Q3-5}
\end{equation}
which implies that the photon number increases (decreases) with the charge
number for $\Delta _{c}\geq \chi q_{s}/\hbar $ ($\Delta _{c}<\chi
q_{s}/\hbar $) in the case of the fixed pumping field. So there should be a
maximal photon number with respect to the change of the charge number, as
demonstrated in Fig. \ref{fig21}(b).

In Fig. 2, the black dotted-dashed and blue solid lines correspond to the
steady-state position and the mean photon number without and with the
approximate condition $n\eta \gg \chi |c_{s}|^{2}$, respectively, as
functions of the charge number. The steady-state position increases
monotonously with the charge number, while the mean photon number is a
pulse-like curve. In the scope of charge number from $30$\ to $55$, there is
a little difference between the exact values and approximate values of both $%
q_{s}$\ and $|c_{s}|^{2}$.\ This deviation results from the fact that the
approximate condition $n\eta \gg \chi |c_{s}|^{2}$\ is not satisfied very
well for the mean photon number $|c_{s}|^{2}\simeq 0.3n\eta /\chi $\ in this
scope. Within the region of $n\leq 40$, the slight difference between the
exact and approximate values implies our assumption $n\eta \gg \chi
|c_{s}|^{2}$\ to be reasonable for the parameters we considered in Fig. \ref%
{fig21}. In this region, both the mean photon number and the MR deformation
are ultra-sensitive to the charge number in a monotonous way, and thereby
the charge number less than $40$ can be fully characterized by the window
width of the OMIT.
\begin{figure}[tbph]
\includegraphics[width=8cm]{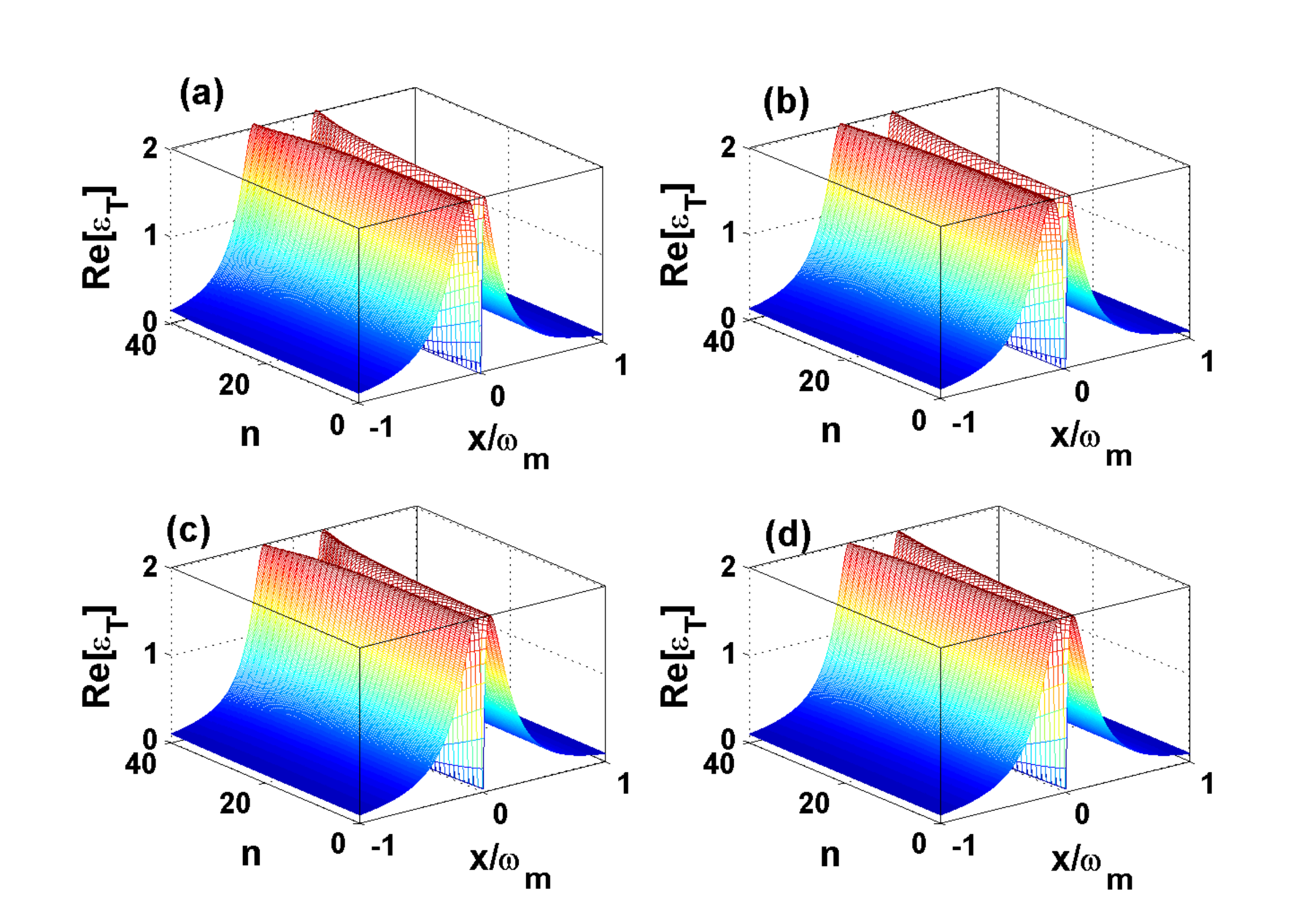}
\caption{(Color online) The real part of the output $Re[\protect\varepsilon %
_{out}]$ (the absorption) versus the charge number $n$ and the detuning $x=%
\protect\delta -\protect\omega _{m}$, where (a) and (b) are the exact values
calculated from Eq. (\protect\ref{Q2-6}). (c) and (d) are the approximate
values using Eq. (\protect\ref{Q2-7}). The parameter values are the same as
in Fig. 2.}
\label{fig 2}
\end{figure}

To show this point more clearly, we have simulated the real part of the
output field using Eq. (\ref{Q2-6}) and Eq. (\ref{Q2-7}) for $n\leq 40$ (see
Fig. \ref{fig 2}). Compared with the exact results, the simplified
expression of the output field Eq. (\ref{Q2-6}) is justified. Moreover, from
Fig. \ref{fig 2}, we see that the absorption vanishes at $x=0$ (i.e., $%
\delta =\omega _{m}$) and the window width of the OMIT increases with the
charge number $n$. So we are able to detect the charge number of a nearby
charged object by the OMIT. In addition, in our case with $n=0$, the values
in the figure can be straightforwardly reduced to those of non-charge case
as in Ref. \cite{pra-81-041803}.
\begin{figure}[tbph]
\includegraphics[width=8cm]{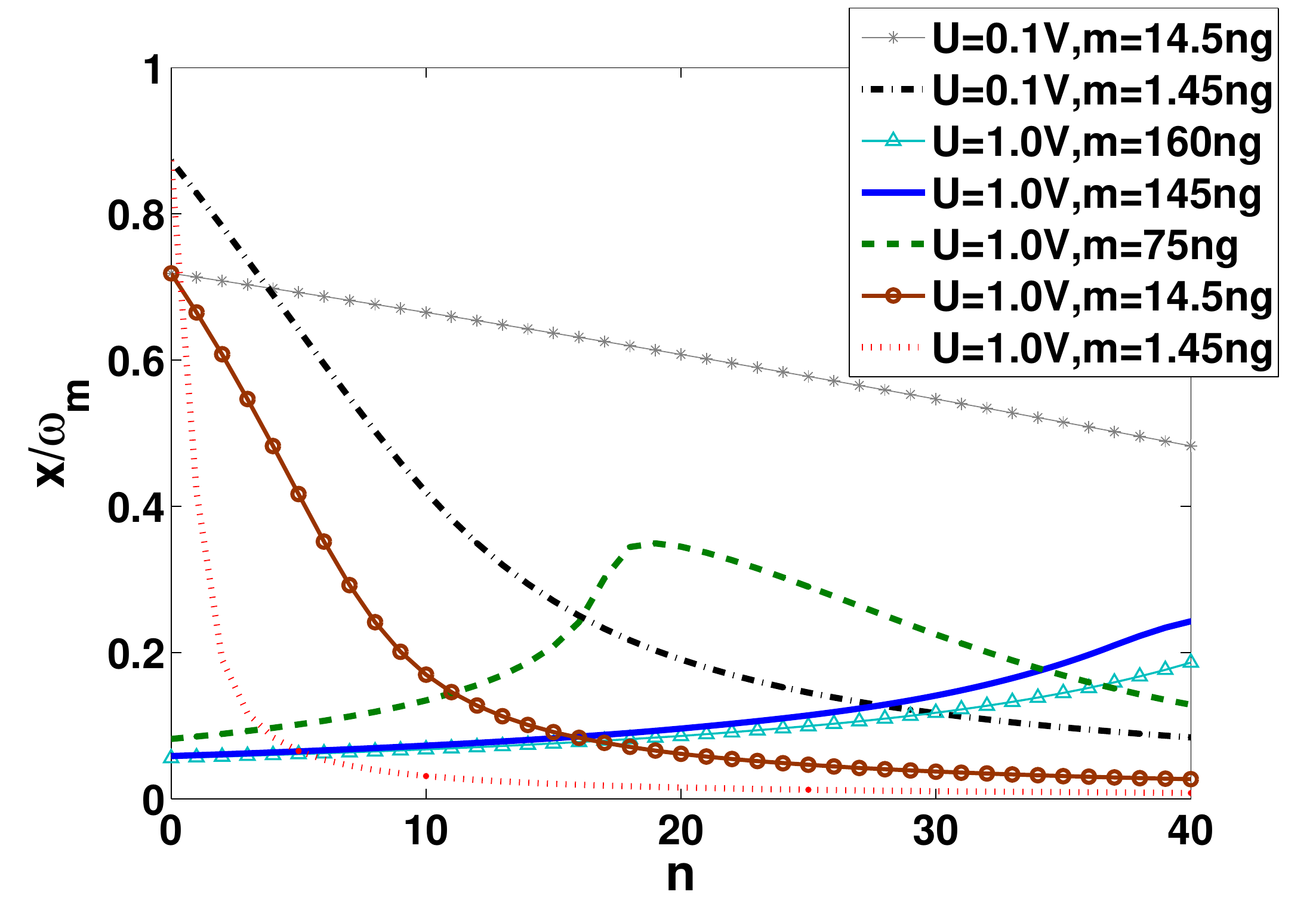}
\caption{(Color online) The detuning $x$ for tuning point versus the charge
number $n$. Except the values in the inset, other parameter values are the
same as in Fig. 2.}
\label{fig4}
\end{figure}

For clarifying the efficiency and the effect of our scheme, we consider a
fixed charge number $n$ in the charged object. There are three tuning points
in the real part of the output field versus the detuning $x=\delta
-\omega_{m}$,
\begin{equation}
\left\{
\begin{array}{l}
x_{\pm }=\pm \sqrt{\dfrac{\sqrt{2}(2\kappa +\gamma _{m})\sqrt{\beta (2\beta
+\kappa \gamma _{m})}-\gamma _{m}(2\beta +\kappa \gamma _{m})}{4\kappa }} \\
x_{0}=0%
\end{array}
\right. ,  \label{Q3-7}
\end{equation}
which can be obtained by solving $d\varepsilon _{T}/dx=0$. Excluding the
trivial case of $x_{0}=0$ and considering the symmetry of $x_{+}$ and $x_{-}$%
, we take the tuning point $x_{+}$ as an example in Fig. \ref{fig4} for
different masses and charges. Since the parameters used in Fig. \ref{fig4}
satisfy $2\beta\gg\kappa\gamma _{m}$, we may reduce Eq. (15) to $x_{+}\sim
\sqrt{\beta}$. Therefore, by changing the MR's effective mass $m$ and the
MR's charge $Q_{MR}$, we present three typically different relationships
between the detuning $x_{+}$ and the charge number $n$: (1) The monotonous
increase (i.e, the gray asterisk and blue solid curves); (2) The monotonous
decrease (i.e., the red dotted and the black dotted-dashed curves); (3) The
curve with a hump in the middle (i.e., the green dashed curve). The curves
indicate that the heavy (light) MR is suitable for detecting large (small)
charge numbers. In addition, although the MR with intermediate mass seems
useless in our scheme, the charge number before the hump ($n<18$) in Fig. 4
can still be used to detect the charge number. Moreover, for the same mass
of the MR, the lower the applied voltage, the less tilting the curve in Fig.
4. As a result, to detect more precisely the tiny charge, e.g., a single
charge, we should increase the voltage to obtain more precise resolution.

\section{Conclusion}

We would like to point out that the analytical solutions to our detection
scheme are based on some approximations ($n\eta \gg \chi |c_{s}|^{2}$ and $%
\delta\sim\omega_{m}$), which have been justified by numerical calculation
with the parameter values we used. It has been shown that the MR with an
effective mass of $1.45$ ng and a voltage of $0.1$ V is more suitable than
others to measure the small charge number (see black dotted-dashed curves in
Fig. \ref{fig4}). Experimentally, the effective mass of a MR as small as 50
pg has been achieved \cite{NatPhys-5-485}. So we may expect to have better
detection with the MRs of such small effective masses.

In addition, from the parameters for the curves in Fig. 4, we can infer the
minimal Coulomb force detected by our scheme, which is $F=k\dfrac{Q_{MR}|e|}{%
r^{2}}=0.88$ nN, much weaker than the minimal electrical gradient force $%
F_{grad}=0.4$ $\mu $N in \cite{prl-104-123604,pra-82-023825}, but larger
than the tiny force $F_{min}=53$ aN detectable in Ref. \cite{njp-14-075015}.
Nevertheless, our scheme focuses on the detection of charge number, rather
than the force. Moreover, the highest sensitivity of the surface charge
density in our scheme is about $1/(0.1r_{0})^{2}\simeq 2.2\times 10^{6}$ cm$%
^{-2}$, which is of the same order as the one ($6.25\times 10^{6}$ cm$^{-2}$%
) in Ref. \cite{Anal. Chem-82-234}. The sensitivity in our case can be
further enhanced by increasing the bias gate voltage or decreasing the mass
of the MR.

In summary, we have demonstrated how to realize precision measurement of
small charge number of the charged object via monitoring the OMIT in
optomechanical system in the presence of the Coulomb interaction between the
charged MR and the object. From the analytical relationship we obtained for
the OMIT window width with the charge number in a small charged body, we
have shown the possibility of detecting few charges (even a single charge)
from the output spectra of the OMIT. The feasibility of our proposal has
been assessed by using currently available parameters, and the Coulomb
attraction under our consideration can be straightforwardly extended to
Coulomb repulsion. We believe that the proposal would be helpful for
exploring quantum behavior in MRs and for precision measurement using OMIT.

\section*{Acknowledgments}

We would like to thank G. S. Agarwal, and Sumei Huang for helpful
discussions. The work is supported by National Fundamental Research Program
of China (Grant Nos. 2012CB922102, 2012CB922104, 2009CB929604 and
2007CB925204), National Natural Science Foundation of China (Grant Nos.
60978009 and 11174027), and Research Funds of Renmin University of China
(Grant No. 10XNL016).

\end{document}